\documentclass[preprint,12pt]{elsarticlemod}

\usepackage{graphicx}
\usepackage{dcolumn}
\usepackage{bm}
\usepackage[english]{babel}
\usepackage[latin1]{inputenc}
\usepackage{graphicx}
\usepackage{epstopdf}
\usepackage{amsfonts}
\usepackage{comment}
\usepackage{url}
\usepackage{color}
\usepackage{epsfig}
\usepackage{mathrsfs}
\usepackage{amsmath,amssymb}
\usepackage{caption}
\usepackage{subcaption}

\usepackage{framed}
\usepackage[svgnames]{xcolor}
\definecolor{shadecolor}{named}{LightGrey}

\newcommand{\lto}[1]{\longrightarrow#1}

\renewcommand{\(}{\left(}
\renewcommand{\)}{\right)}
\renewcommand{\[}{\left[}
\renewcommand{\]}{\right]}

\usepackage{amssymb}

\begin{document}

\selectlanguage{english}

\begin{frontmatter}



\title{Random Walk Centrality in Interconnected Multilayer Networks}


\author{Albert Sol\'e-Ribalta}
\author{Manlio De Domenico}
\author{Sergio G\'omez}
\author{Alex Arenas*}
\cortext[cor1]{email: alexandre.arenas@urv.cat}

\address{Departament d'Enginyeria Inform\`atica i Matem\`atiques, Universitat Rovira i Virgili, 43007 Tarragona, Spain}

\begin{abstract}
Real-world complex systems exhibit multiple levels of relationships. In many cases they require to be modeled as interconnected multilayer networks, characterizing interactions of several types simultaneously.
It is of crucial importance in many fields, from economics to biology and from urban planning to social sciences, to identify the most (or the less) influent nodes in a network using centrality measures.
However, defining the centrality of actors in interconnected complex networks is not trivial. In this paper, we rely on the tensorial formalism recently proposed to characterize and investigate this kind of complex topologies, and extend
two well known random walk centrality measures, the random walk betweenness and closeness centrality, to interconnected multilayer networks.
For each of the measures we provide analytical expressions that completely agree with numerically results.
\end{abstract}


\end{frontmatter}



\section{Introduction}

It is common practice in many studies involving networks to assume that nodes are connected by a single type of edge that encapsulates all relations between them. In a myriad of applications this assumption oversimplifies the complexity of the system, leading to inaccurate or wrong results. Examples can be found in temporal networks, where neglecting time-dependence washes out the memory of sequences of human contacts in transmission of diseases \cite{holme2012}, in co-authorship networks, where neglecting the existence of multiple relationships between actors might alter the topology which may lead to misestimating crucial node's properties \cite{freeman1979centrality, jeong2001lethality, guimera2002optimal, guimera2005worldwide, barthelemy2004betweenness, nicosia2012controlling} or in transportation networks where the multilayer topology must be considered to accurately model the dynamics to \textit{a posteriori} predict congested locations \cite{deDomenico2015ranking}.

\begin{figure*}[!ht]
    \begin{center}
    	\begin{tabular}{lll}
                {\bf A} & {\bf B} & {\bf C}
                \\
                \includegraphics[width=0.3\textwidth]{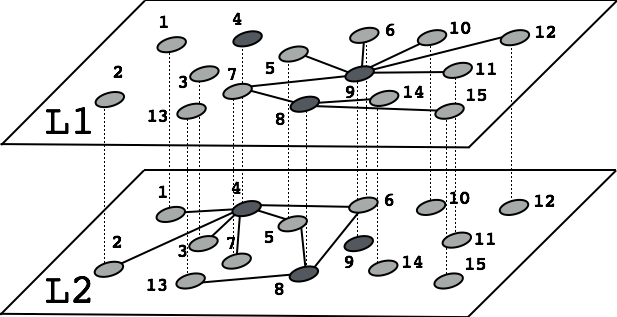}
                &
                \includegraphics[width=0.3\textwidth]{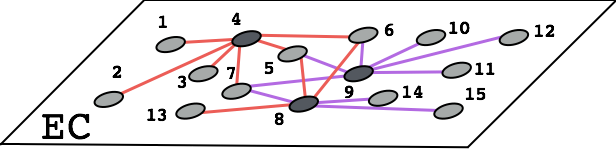}
                &
                \includegraphics[width=0.3\textwidth]{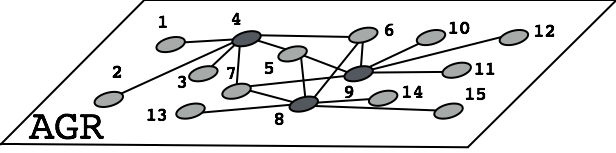}
        \end{tabular}
    \end{center}
        \caption{{\bf A}, an interconnected multilayer network representing the same actors exhibiting different relationships on different levels. The cost to move from one layer to the other is represented by dotted vertical lines. {\bf B}, edge-colored graph representing the same actors with the same relations in {\bf A} with two different types of interactions (solid and dashed edges). In this case the representation does not allow modeling the cost to move between layers. {\bf C}, classical approach of representing the different types of relations using an aggregated network. The network represents the same actor and relation in {\bf A} and {\bf B} but disregarding the type of relation.}
        \label{fig:mplex-vs-edgecolor}
\end{figure*}

Historically, the term \emph{multiplex} was coined to indicate the presence of more than one relationship between the same actors of a social network \cite{padgett1993robust}. This type of network is well understood in terms of ``coloring'' (or labeling) the edges corresponding to interactions of different nature. For instance, in a social network the same individual might have connections to other individuals based on financial interests (e.g., color red) and connections with the same or different individuals based on friendship (e.g., color blue).
In other real-world systems, like the transportation network of a city, the same geographical position can be part, for instance, of the network of subway or the network of bus routes, simultaneously. In this specific case, an edge-colored graph would not capture the full structure of the network, since information about the cost to \emph{move} from the subway network to the bus route is missing. This cost can be economic or might account for the time required to physically commute between the two layers. It is in this cases where an the interconnected multilayer network provides a better representation of the system. Figure~\ref{fig:mplex-vs-edgecolor} shows an illustration of an interconnected multilayer (Fig.\,\ref{fig:mplex-vs-edgecolor}~A) and the classical representation with an aggregated network (Fig.\,\ref{fig:mplex-vs-edgecolor}~C). It is evident that a simple projection of the former -- mathematically equivalent to sum up the corresponding adjacency matrices of the individual layers -- would provide a network where the information about the relation type is lost. On the other hand, an edge-colored graph (Fig.\,\ref{fig:mplex-vs-edgecolor}~B) can not account for interconnections. For further details about the classification of such multilayer networks we refer to \cite{kivela2013multilayer} and references therein. In the rest of the paper interconnected multilayer networks will be referred in short as multilayer networks.

The remainder of this paper is organized as follows. In Sec.\,\ref{sec:tensors} we briefly describe the tensorial notation, defined in \cite{dedomenico2013mathematical}, adopted overall the paper. In Sec.\,\ref{sec:centrality} we capitalize on this notation to extend some random walk centrality descriptors, well known in the case of single layer networks, to interconnected multilayer networks. Finally, we discuss our findings in Sec.\,\ref{sec:conclusion}.

\section{Tensorial notation}\label{sec:tensors}

Edge-colored graphs can be represented by a set of adjacency matrices \cite{cardillo2013emergence,nicosia2013growing,Bianconi2013Statistical,battiston2013metrics}. However, standard matrices, used to represent networks, are limited in the complexity of the relationships that they can capture, i.e., they do not represent a suitable framework in the case of multilayer networks. This is the case of multiple types of relationships -- that can also change in time -- between nodes. Such a level of complexity can be characterized by considering tensors and algebras of higher order \cite{dedomenico2013mathematical}.

A great advantage of tensor formalism developed in \cite{dedomenico2013mathematical} relies on its compactness. An adjacency tensor can be written using a more compact notation that is very useful for the generalization of network descriptors to multilayer networks. In this notation, a row vector $\mathbf{a}\in\mathbb{R}^{N}$ is given by a covariant vector $a_{\alpha}$ ($\alpha=1,2,\ldots,N$), and the corresponding contravariant vector $a^{\alpha}$ (i.e., its dual vector) is a column vector in Euclidean space. A canonical vector is assigned to each node and the corresponding interconnected multi-layer network is represented by a mixed rank-4 adjacency tensor.

However, in the majority of applications, it is not necessary to perform calculations using canonical vectors and tensors explicitly. In this cases, a classical single-layer network can be represented by a rank-2 mixed adjacency tensor $W^{\alpha}_{\beta}$ \cite{dedomenico2013mathematical}, where the layer information is disregarded. But, in general, systems may exhibit several types of relationships between pairs of nodes and a more general system represented as a multilayer object -- in which each type of relationship is represented within a single \emph{layer} $\alpha$ ($\alpha=1,2,\ldots,L$) of the network -- is required\footnote{To avoid confusion, in the following we refer to nodes with Latin letters and to layers with Greek letters, allowing us to distinguish indices that correspond to nodes from those that correspond to layers in tensorial equations.}. In these cases, we use an \emph{intra-layer adjacency tensor} for the $2^{\text{nd}}$-order tensor $W^{i}_{j}(\alpha)$ that indicates the relationships between nodes within the \emph{same} layer $\alpha$ and the $2^{\text{nd}}$-order \emph{inter-layer adjacency tensor} $C^{i}_{j}(\alpha\beta)$ to encode information about relationships that incorporate multiple layers.

It has been shown that the mathematical object accounting for the whole interconnected multilayer structure is given by a $4^{\text{th}}$-order (i.e., rank-4) \emph{multilayer adjacency tensor} $M^{i\alpha}_{j\beta}$. This tensor might be simply thought as a higher-order matrix with four indices. It is the direct generalization of the adjacency matrix in the case of single layer networks and encodes the intensity of the relationship (which may not be symmetric) between a node $i$ in layer $\alpha$ and a node $j$ in layer $\beta$ \cite{dedomenico2013mathematical}.

To reduce the notational complexity in the tensorial equations the Einstein summation convention is adopted.
It is applied to repeated indices in operations that involve tensors. For example, we use this convention in the left-hand sides of the following equations:
\begin{align}
	A^{i}_{i} = \sum_{i=1}^{N}A^{i}_{i}\,,\quad
	A^{i}_{j}B^{j}_{i} =\sum_{i=1}^{N}\sum_{j=1}^{N}A^{i}_{j}B^{i}_{j}\,,\nonumber\\
	A^{i\alpha}_{j\beta}B^{k\beta}_{i\gamma} =\sum_{i=1}^{N}\sum_{\beta=1}^{L}A^{i\alpha}_{j\beta}B^{k\beta}_{i\gamma}\,,\nonumber
\end{align}
whose right-hand sides include the summation signs explicitly.  It is straightforward to use this convention for the product of any number of tensors of any order. In the following, we will use the $t$-th power of rank-4 tensors, defined by multiple tensor multiplications:
\begin{equation}
(A^t)^{i\alpha}_{j\beta} = (A)^{i\alpha}_{j_1\beta_1}(A)^{j_1\beta_1}_{j_2\beta_2}\dots(A)^{j_{t-1}\beta_{t-1}}_{j\beta}
\end{equation}

Repeated indices, such that one index is a subscript and the other is a superscript, is equivalent to perform a tensorial operation known as a \emph{contraction}. Moreover, one should be very careful in performing tensorial calculations. For instance, using traditional notation the product $a^{i}b^{j}$ would be a number, i.e., the product of the components of two vectors. However, in our formulation, the same calculation denotes a Kronecker product between two vectors, resulting in a rank-2 tensor, i.e., a matrix.


\section{Random walk centrality measures in multilayer networks}\label{sec:centrality}

In practical applications one is often interested in assigning a global measure of importance to each node. If the system we deal with contains several types of relations between actors we expect that the measures, in some way, consider the importance obtained from the different layers. A simple choice could be to combine the centrality of the nodes -- obtained from the different layers independently -- according to some heuristic choice. This is a viable solution when there is no interconnection between layers, i.e., in the case of edge-colored graphs \cite{Sola2013Centrality,pagerank2013}. However, the main drawback of this approach is that it depends on the choice of the heuristics and thus might not evaluate the actual importance of nodes. Our approach accounts for the higher level of complexity of such systems without relying on external assumptions and naturally extends the well-known centrality measures adopted for several decades in the case of single layer networks.

A random walk is the simplest dynamical process that can occur on a network, and random walks can be used to approximate other types of diffusion processes\cite{chung1997,Newman2010Book}. Random walks on networks \cite{chung1997,noh2004random,Newman2010Book} have attracted considerable interest because they are both important and easy to interpret. They have yielded important insights on a huge variety of applications and can be studied analytically. For example, random walks have been used to rank Web pages \cite{pagerank1998} and sports teams \cite{callaghan2007}, optimize searches \cite{viswanathan1999optimizing}, investigate the efficiency of network navigation \cite{yang2005exploring,da2007exploring}, characterize cyclic structures in networks \cite{rozenfeld2005statistics}, and coarse-grain networks to highlight meso-scale features such as community structure \cite{gfeller2007spectral,rosvall2007information,lambiotte2008}. Another interesting application of random walks is to calculate the centrality of actors in complex networks when there is no knowledge about the full network topology but only local information is available. In such cases, centrality descriptors based on shortest-paths, e.g., betweenness and closeness centrality, should be substituted by centrality notions based on random walks \cite{noh2004random,newman2005measure}. In the following we extend these measures to multilayer networks.

\begin{figure}[!t]
	\centering
	  \includegraphics[width=8cm]{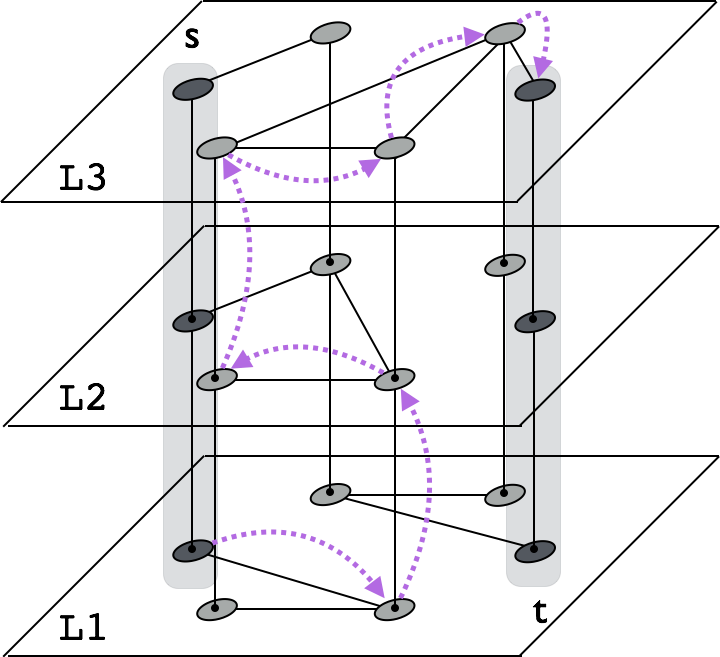}
	\caption{Schematic of a walk (dotted trajectories) between two individuals $s$ and $t$ using a multilayer network. A walker can jump between nodes within the same layer, or it might switch to another layer. This illustration evidences how multilayer structure allows a walker to move between nodes that belong to different (disconnected) components on a given layer (L1).}
        \label{fig:mplex-rw}
\end{figure}

First of all, we define a discrete-time random walk, between two individuals $s$ and $t$, $s \rightarrow t$, on a multilayer network consisting of $L$ layers and $N$ nodes per layer, as a random sequence of nodes which starts from node $s$ in any layer and finish in node $t$ in any layer where each edge's endpoints are the preceding and following vertices in the sequence. The reasoning behind this definition is that the different node replicas in the different layers correspond to the same individual and so anything traveling between them is independent on the starting and ending layer. Fig.~\ref{fig:mplex-rw} shows and example of a random walk between two nodes in a multilayer network where it is evident the introduction of non-trivial effects because of the presence of inter-layer connections that affects its navigation in the networked system \cite{dedomenico2013random}.

\emph{Random walk occupation centrality.} Let $T^{i\alpha}_{j\beta}$ denote the tensor of transition probabilities for jumping between pairs of nodes and switching between pairs of layers. Covariant indexes $i\alpha$ indicate source node and layer and contravariant indexes $j\beta$ destination node and layer. Similarly to the single layer case, the sum of the probabilities for each outgoing edges of any node adds one. That is, $u_{i\alpha} = T^{i\alpha}_{j\beta} u^{j\beta}$ where $u_{i\alpha}$ and $u^{j\beta}$ are the 1-row vector and 1-column vector respectively. In addition, let $p_{i\alpha}(t)$ be the time-dependent tensor that gives the probability to find a walker at a particular node in a particular layer. Hence, the covariant master equation that governs the discrete-time evolution of the probability from time $t$ to time $t+1$ is $p_{j\beta}(t+1)=T^{i\alpha}_{j\beta}p_{i\alpha}(t)$.  The steady-state solution of this equation, i.e., for $t\lto\infty$, is given by $\Pi_{i\alpha}$ which quantifies the probability to find a walker in the node $i$ of layer $\alpha$. In the case of single layer networks, the steady-state solution can be obtained by calculating the leading eigenvector corresponding to the unitary eigenvalue. Similarly, in the case of multilayer networks, the solution can be obtained by calculating the leading \emph{eigentensor}, solution of the higher-order eigenvalue problem
\begin{eqnarray}
T^{i\alpha}_{j\beta}\Pi_{i\alpha}=\lambda\Pi_{j\beta}.
\end{eqnarray}
We refer to Appendix\,\ref{sec:eigentensor} for the mathematical details to solve this problem. The probability $\Pi_{j\beta}$, defined as \emph{random walk occupation centrality}, accounts for the full interconnected structure of the multilayer network. Although different exploration strategies can be adopted to define the transition tensor $T^{i\alpha}_{j\beta}$ walk in a multilayer network \cite{dedomenico2013random}, here we only focus on the natural extension of well-known random walks in single layer networks \cite{noh2004random}. In this process, the walker in node $i$ and layer $\alpha$ might jump to one of its neighbors $j\neq i$ -- within the same layer -- or might switch to its counterpart $i$ in a different interconnected layer $\beta\neq\alpha$ with equal probability. That is, the inter-layer connection is treated as an edge that can be chosen randomly among all outgoing edges of the node. In the more general case of weighted networks, the jumping probability is usually defined proportional to the weight of the edges. Let us indicate with $s_{i\alpha}$ the strength of node $i$ in layer $\alpha$, including the inter-layer connections. The multi-strength vector, whose components indicate the strength of each node accounting for the full multilayer structure, is given by summing up its strengths across all layers, i.e., by $S_{i}=s_{i\alpha}u^{\alpha}$. We indicate with $D^{i\alpha}_{j\beta}$ the strength tensor whose entries are all zeros, except for $i=j$ and $\alpha=\beta$ where the entries are given by $s_{i\alpha}$. This tensor represents the multilayer extension of the well-known diagonal strength matrix in the case of single layer networks. Therefore, the transition tensor is given by $T^{i\alpha}_{j\beta}=M^{k\gamma}_{j\beta}\tilde{D}^{i\alpha}_{k\gamma}$, where $\tilde{D}^{i\alpha}_{j\beta}$ is the tensor whose entries are the inverse\footnote{It is worth remarking that, in general, this is different from the inverse of a tensor $A^{i\alpha}_{j\beta}$, that is defined as the tensor $B^{i\alpha}_{j\beta}$ such that $A^{i\alpha}_{k\gamma}B_{j\beta}^{k\gamma}=\delta^{i\alpha}_{j\beta}$, where $\delta^{i\alpha}_{j\beta}=\delta^{i}_{j}\delta^{\alpha}_{\beta}$.} of the non-zero entries of the strength tensor and $M^{k\gamma}_{j\beta}$ is the weighted adjacency tensor. For this classical random walk, it can be easily shown that $\Pi_{i\alpha}\propto s_{i\alpha}$ \cite{dedomenico2013random}.

This centrality, as others in the rest of the paper, assigns a measure of importance to each node in each layer, accounting for the full-interconnected structure of the multilayer network. However, in practical applications one is often interested in assigning a global measure of importance to each node, aggregating the information obtained from the different layers. The choice of the aggregation method is in general not trivial and it strongly influences the final estimation and might lead to wrong results.

However, this is not case for the occupation probability. Since the centrality $\Pi_{i\alpha}$ is calculated accounting for the full interconnected structure of the whole system we do not require any arbitrary combination of the information from different layers. In our framework, the most intuitive type of aggregation, i.e., summing up over layers, represents the unique and correct choice. Let $\pi_{i}=\Pi_{i\alpha}u^{\alpha}$ be the random walk centrality measure obtained by aggregating over the layers. Here, $\pi_{i}$ indicates the probability of finding the walker in node $i$, regardless of the layer. It is worth noting that this probability, as well as in single layer networks is, is proportional to $s_{i\alpha}u^{\alpha}$, i.e., the multi-strength of node $i$. Therefore, in this specific case, the computation of the centrality by means of the aggregated network would provide the same result of the calculation accounting for the multilayer structure, if inter-layer edges are mapped to self-loops. Unfortunately, this is not the case for the other centrality measures discussed in the rest of this study, where calculating the diagnostics from the aggregate might lead to wrong conclusions.

A measure related to random walk occupation centrality is the Page Rank \cite{brin1998} that has been recently extended to interconnected networks \cite{deDomenico2015ranking}. In fact, the Page Rank centrality can be seen as the steady-state solution of the random walk master equation governed by the transition tensor $R^{i\alpha}_{j\beta}$, where the walker jumps to a neighbor with rate $r$ and teleport to any other node in the network with rate $1-r$. This rank-4 tensor is given by
\begin{equation}\label{RWPageRank}
	R^{i\alpha}_{j\beta} = rT^{i\alpha}_{j\beta} + \frac{(1-r)}{NL}u^{i\alpha}_{j\beta},
\end{equation}
where $u^{i\alpha}_{j\beta}$ is the rank-4 tensor with all components equal to 1.
The steady-state solution of the master equation corresponding to this transition tensor provides the Page Rank centrality for multilayer networks.

\vspace{0.25truecm}\emph{Random walk betweenness centrality.} The betweenness is a measure of network centrality that instead of accounting for topological centrality accounts for the importance of nodes in terms of dynamical processes that run over the network. In particular, the betweenness measures to which extent a node lies in the path between any two other nodes \cite{newman2005measure}. One can think of packets traveling in internet, in this case the betweenness measures the influence of nodes in the controling of information. The most common betweenness is the shortest path betweenness \cite{Freeman77centralityin} where the centrality of a node $j$ is relative to the number of shortest paths, for any pair $(o,d)$ of \emph{origin} and \emph{destination} nodes, that pass through $j$. However, in real networks, entities (rumors, messages or packets over the Internet) that travel the network do not always take the shortest path \cite{Freeman91centralityin, Stephenson19891}. Consider, for instance, rumors that can be wandering around the network or packets trying to avoid overloaded routers. In such cases, the shortest path betweenness is not always a good proxy for the centrality of nodes. For these scenarios the random walk betweenness of a node $j$ is defined as the amount of random walks between any pair $(o,d)$ of nodes that pass through $j$ \cite{newman2005measure}.

To analytically compute the number of random walks visiting a particular node, it is often convenient to use the concept of absorbing random walk, where the absorbing state is selected to be the destination node $d$ \cite{newman2005measure,Newman2010Book}. To extend this concept to the case of interconnected multilayer networks, we consider random walks that begin, pass and end in nodes in different layers while accounting for the existence of several replicas of the same node.

Specifically, to extend the concept of random walks to interconnected networks, we define the absorbing transition tensor on a particular node $d$ by
\begin{eqnarray} \label{eq:absTranTensor}
	\(T_{[d]}\)^{i\alpha}_{j\beta} = \left\{
		\begin{array}{l l}
		0 & \quad j=d\\
		T^{i\alpha}_{j\beta} & \quad \mbox{else}
	\end{array} \right.,
\end{eqnarray}
Random walkers governed by this transition tensor will vanish once they arrive to any absorbing state \cite{Newman2010Book}. Note that $T_{[d]}$ has one absorbing state for each replica of node $d$ in different layers. It can be shown (see Appendix \ref{ap:timesOnNodeRWB}) that the average number of times a random walk (with origin in node $o$ in layer $\sigma$ and destination $d$ independently of the layer) will pass by a node $j$ in layer $\beta$, regardless of the time step, is given by
\begin{eqnarray}
\({\tau_{[d]}}\)^{i\alpha}_{j\beta} = \[\(\delta - T_{[d]}\)^{-1}\]^{i\alpha}_{j\beta},
\end{eqnarray}
where $\delta^{i\alpha}_{j\beta}=\delta^{i}_{j}\delta^{\alpha}_{\beta}$ and $\delta$ is the the Kronecker delta. Note that the average number of times that the walk will visit node $j$ still depends on the layer where $j$ is located and on the originating layer $\sigma$. Since we are interested on node properties, regardless of the layer, we average over all possible starting layers $\sigma$ and aggregate the walks that pass through $j$ in the different layers,
\begin{eqnarray}
	\(\tau_{[d]}\)_{j}^{o} &=& \frac{1}{L}\({\tau_{[d]}}\)^{o\sigma}_{j\beta}u^{\beta}u_{\sigma}.
\end{eqnarray}
The overall centrality vector is obtained by averaging over all possible origins and destinations:
\begin{eqnarray}\label{eq:anRWBetweenness}
\tau_{j}  = \frac{1}{N(N-1)}\sum\limits_{d = 1}^{N} \(\tau_{[d]}\)_{j}^{o}u_{o}.
\end{eqnarray}

\begin{figure*}[!t]
    \begin{center}
    	\begin{tabular}{lll}
                {\bf A} & {\bf B} & {\bf C}
                \\
                \includegraphics[width=0.3\textwidth]{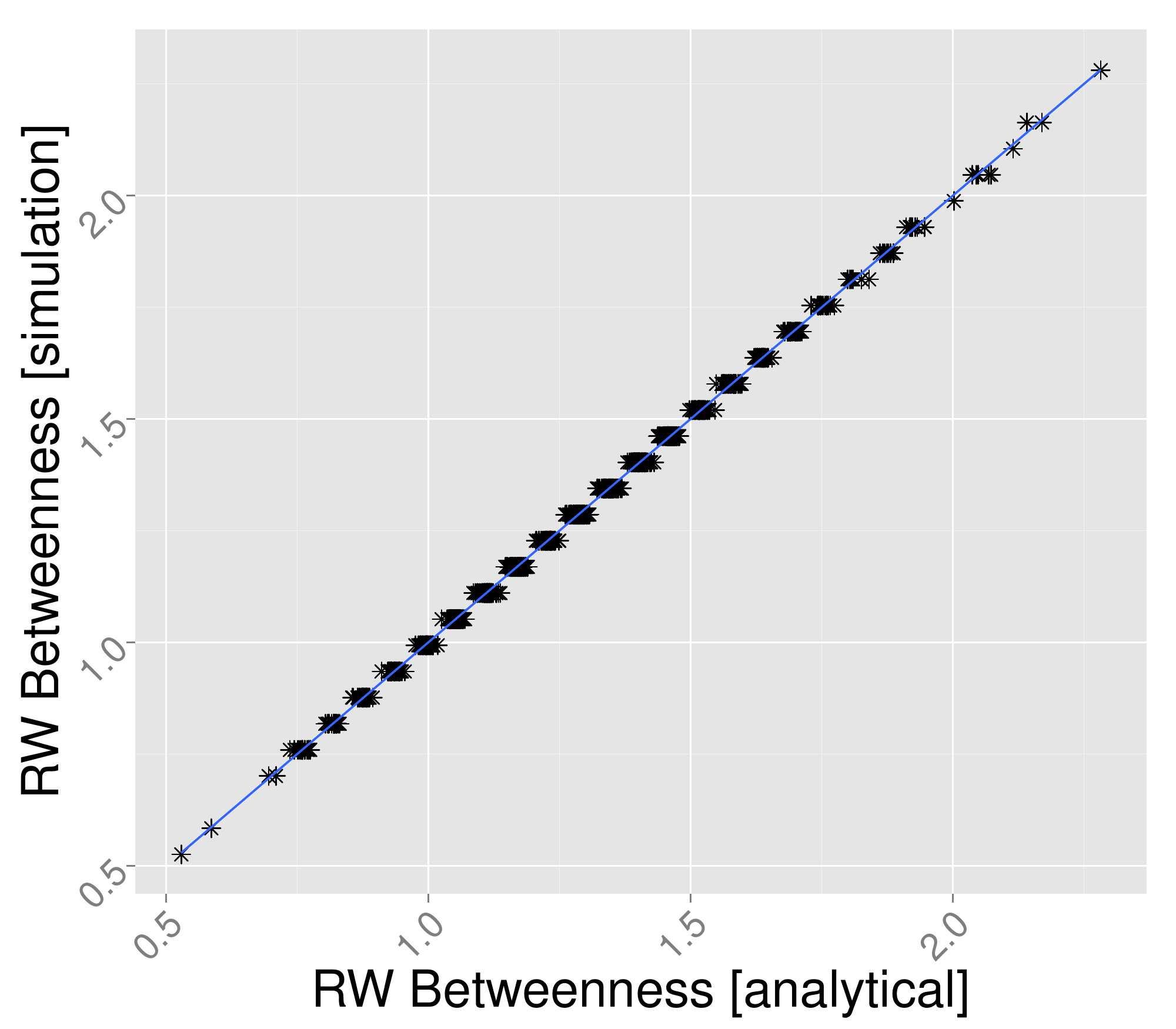}
                &
                \includegraphics[width=0.3\textwidth]{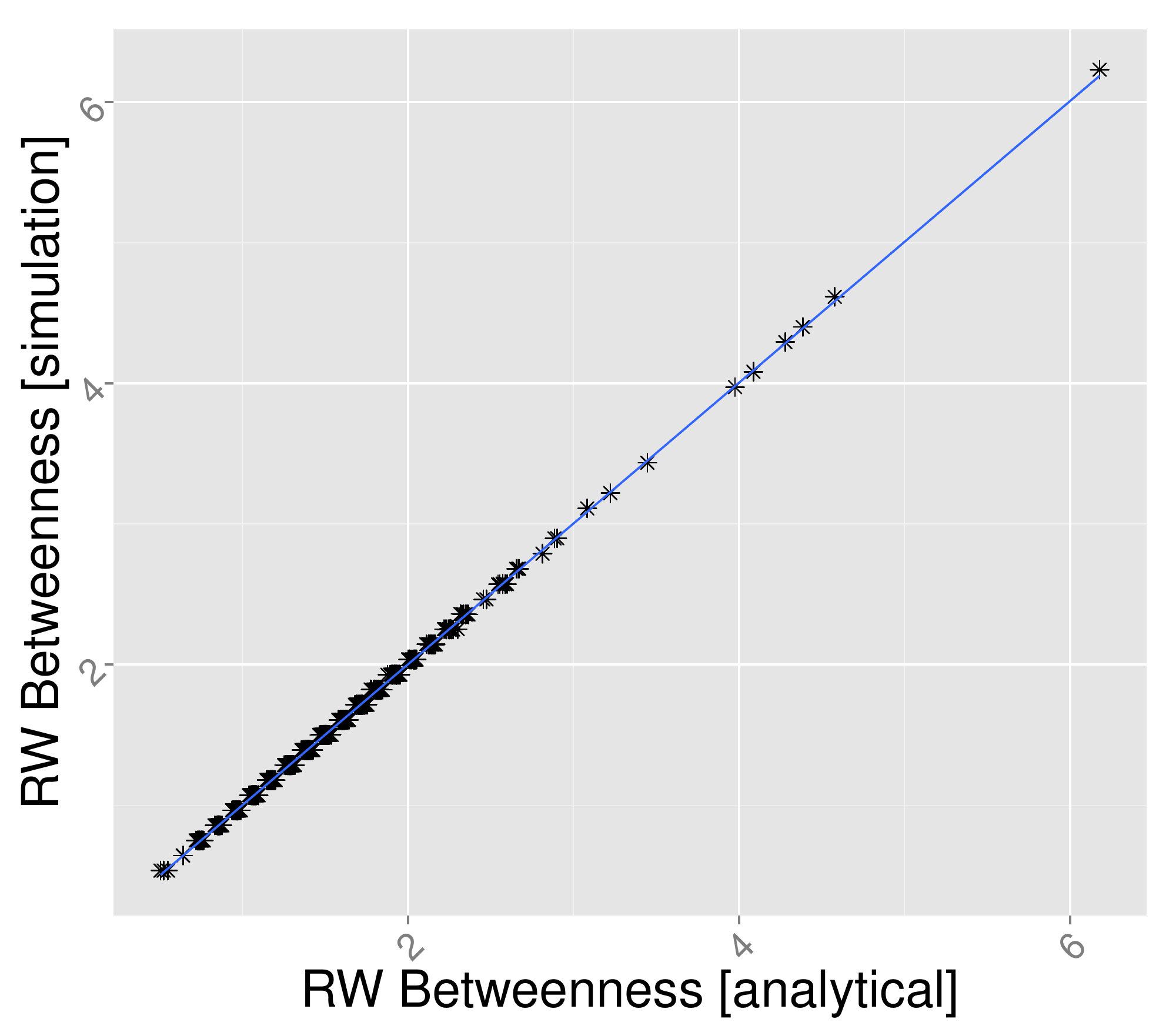}
                &
                \includegraphics[width=0.3\textwidth]{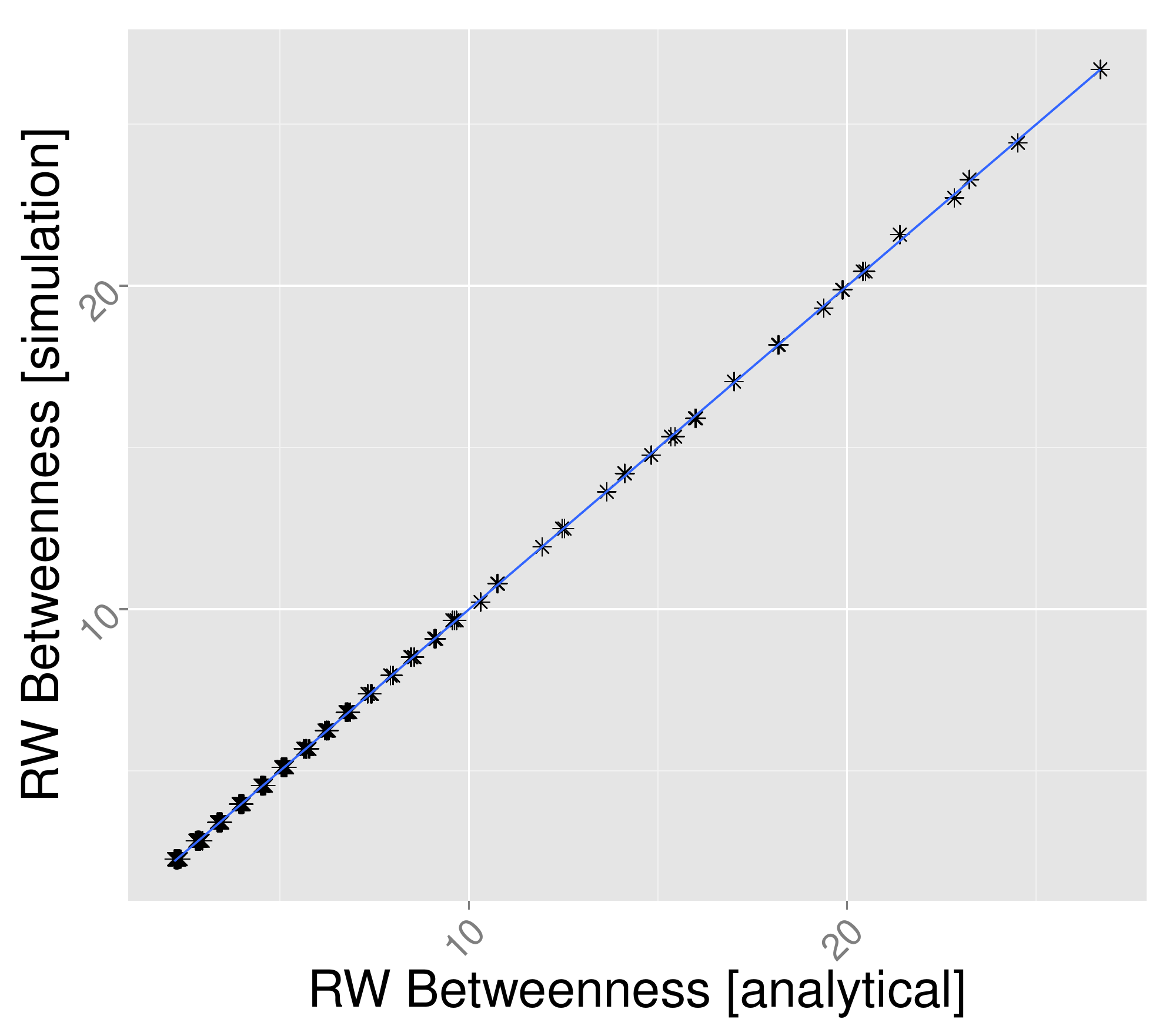}
        \end{tabular}
    \end{center}
        \caption{Comparison of the random walk betweenness centrality obtained by simulation and by our analytical approach for different multilayer network topologies. Each multilayer network is composed of two layers with 1000 nodes per layer. {\bf A}, results on a multilayer network with two Erd{\H{o}}s-R\'{e}nyi networks as layers. {\bf B}, results on a multilayer network with one Erd{\H{o}}s-R\'{e}nyi network and one Barab\`{a}si-Albert network as layers. {\bf C}, results on a multilayer network with two Barab\`{a}si-Albert networks as layers.}
        \label{fig:evaluationBetweeness}
\end{figure*}

The comparison between the values of $\tau_{i}$ obtained from simulations and theoretical predictions are shown in Fig.\,\ref{fig:evaluationBetweeness}. As expected, the results are in excellent agreement. It is worth remarking that the equivalence holds regardless of the number of nodes in the network, the topology and the number of layers.

\vspace{0.25truecm}\emph{Random walk closeness centrality.} The distance between two nodes in a network is given by shortest-path which separates them. The farness of an individual is given by the sum of all geodesics from that node to any other node. In general, the inverse of this farness provides a measure of the \emph{closeness} of the node. Such a diagnostic is related to how fast information is expected to spread from a given actor to the others in the network. A variant of the closeness when random walks are considered is given by the random walk closeness centrality. In the case of single layer networks, it has been introduced to quantify how central a node is located regarding its potential to receive information randomly diffusing over the network \cite{noh2004random}.

We define the random walk closeness centrality of a node $i$ in a multilayer network as the inverse of the average number of steps that a random walker, starting from any other node in the multilayer network, requires to reach $i$ for the first time. The computation of the closeness centrality is generally based on the mean first-passage time (MFPT), that is defined as the average number of steps to reach a node $d$, starting from a given node $s$. The MFPT matrix can be computed analytically by means of Kemeny-Snell fundamental matrix $Z$ \cite{lovasz1993random,zhang2011mean} or by means of absorbing random walks \cite{kemeny1960finite,Newman2010Book}. In this study, we adopt the second approach as for the calculation of random walk betweenness centrality. The following calculation involve the use of the transition tensor $T$ governing random walks over multilayer networks and the corresponding absorbing transition tensor $T_{[d]}$. Hence, the tensor
\begin{equation}
p_{j\beta}^{o\sigma}(t)= \(T^t_{[d]}\)_{j\beta}^{o\sigma}
\end{equation}
indicates the probability of visiting node $j$ in layer $\beta$, after $t$ time steps, considering that the walk originated in node $o$ in layer $\sigma$. This transition tensor is absorbing on node $d$ regardless of the layer and, consequently, any walker reaching an absorbing state will vanish, i.e., $p_{d\beta}^{o\sigma}(t)=0$ for any $\beta$ and $t$. The probability that the walker is absorbed in some node $d$ at a time $h$ equal or smaller than $t$, regardless of the layer, is given by
\begin{equation}
\(q_{[d]}\)^{o\sigma}(t)= u^{o\sigma}-\(T^t_{[d]}\)_{j\beta}^{o\sigma}u^{j\beta} \label{eq:probOfBeingAbsorbed}.
\end{equation}
Note that we have a rank-2 tensor $q$ for each choice of $d$ and we put in evidence this dependence by means of $[d]$. From each tensor $q$ we can calculate the probability that the first passage time for node $d$ is exactly $t$ by
\begin{eqnarray}
\(q_{[d]}\)^{o\sigma}(h = t) &=& \(q_{[d]}\)^{o\sigma}(t)-\(q_{[d]}\)^{o\sigma}(t-1) \nonumber \\
&=& \[\(T^t_{[d]}\)-\(T^{t-1}_{[d]}\)\]_{j\beta}^{o\sigma}u^{j\beta}. \label{eq:pdf_closeness}
\end{eqnarray}
Considering the walk starts from node $o$ in layer $\sigma$, each tensor encoding the mean first passage time to node $d$ is obtained from Eq.\,(\ref{eq:pdf_closeness}) as
\begin{eqnarray}
\(H_{[d]}\)^{o\sigma} = \sum\limits_{t=0}^{\infty} t \(q_{[d]}\)^{o\sigma}(h=t)  = \[\(\delta-T^{t}_{[d]}\)^{-1}\]_{j\beta}^{o\sigma}u^{j\beta}. \label{eq:meanTau}
\end{eqnarray}

The geometric series in Eq.\,(\ref{eq:meanTau}) converges since the maximum eigenvalue of $T_{[d]}$ is strictly smaller than one, and the sum can be calculated exploiting the self-similarity of the series. Note that the mean first passage time to $d$ still depends on the origin of the walk, i.e., node $o$ in layer $\sigma$.

\begin{figure*}[!t]
    \begin{center}
    	\begin{tabular}{lll}
                {\bf A} & {\bf B} & {\bf C}
                \\
                \includegraphics[width=0.3\textwidth]{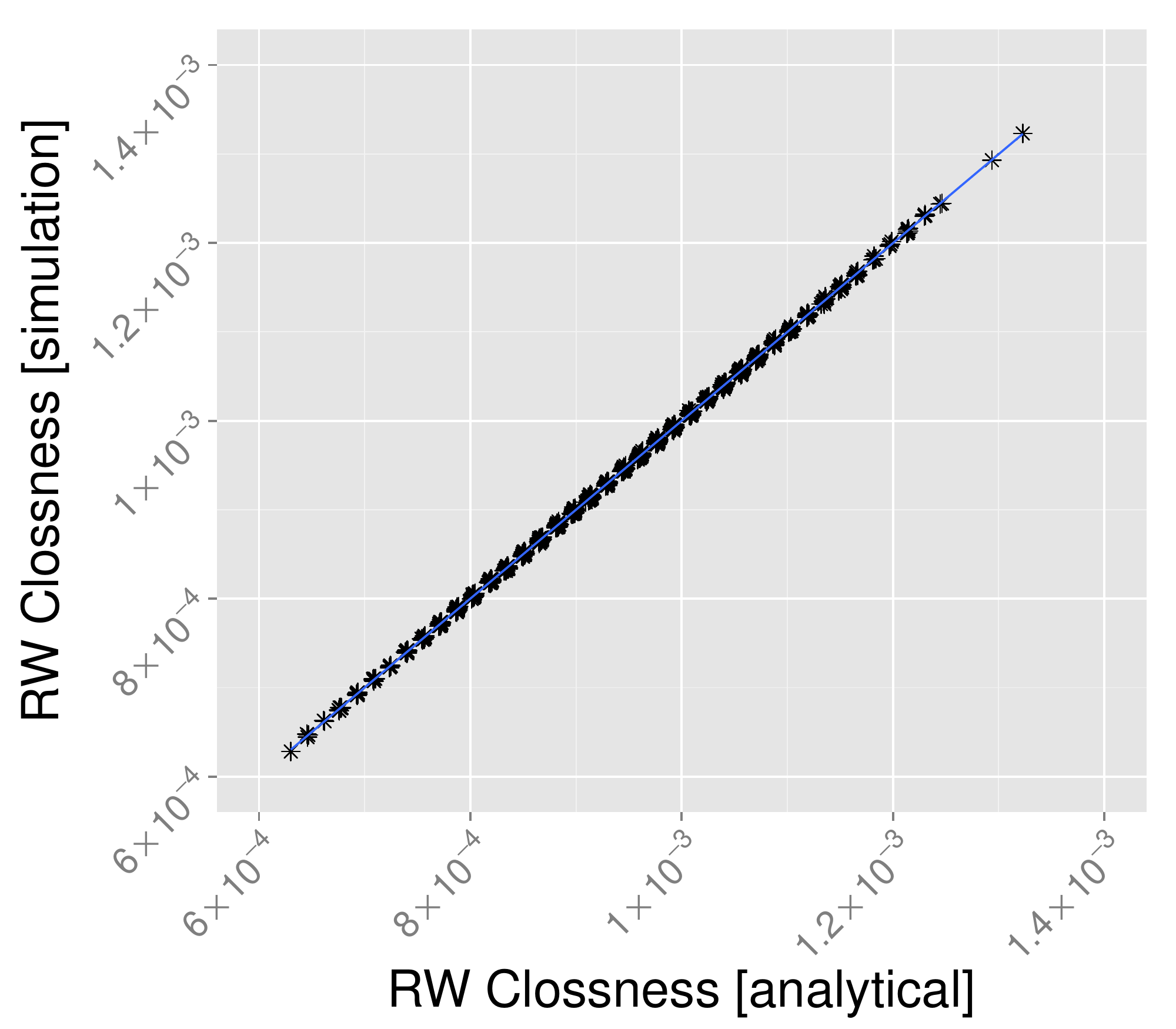}
                &
                \includegraphics[width=0.3\textwidth]{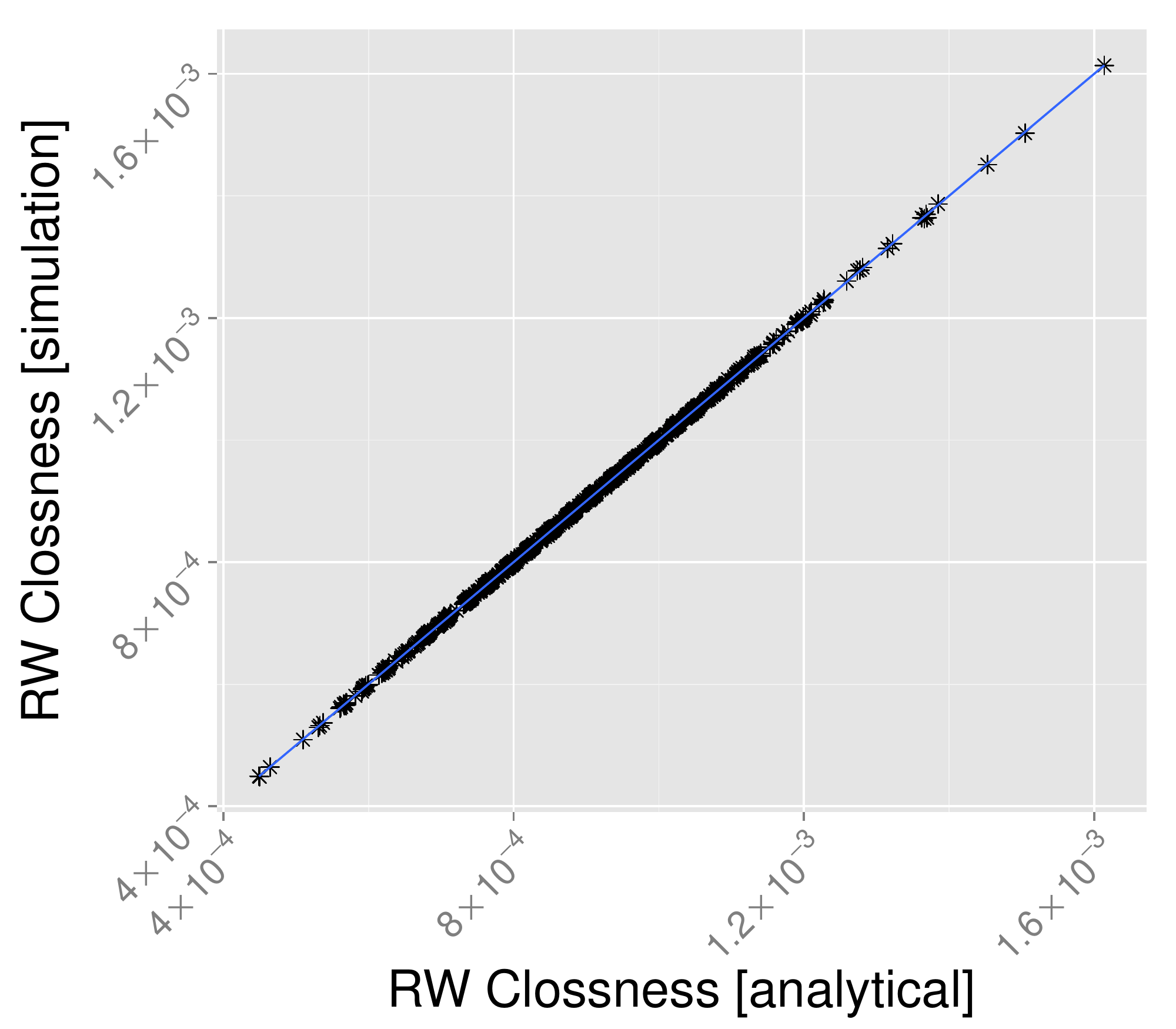}
                &
                \includegraphics[width=0.3\textwidth]{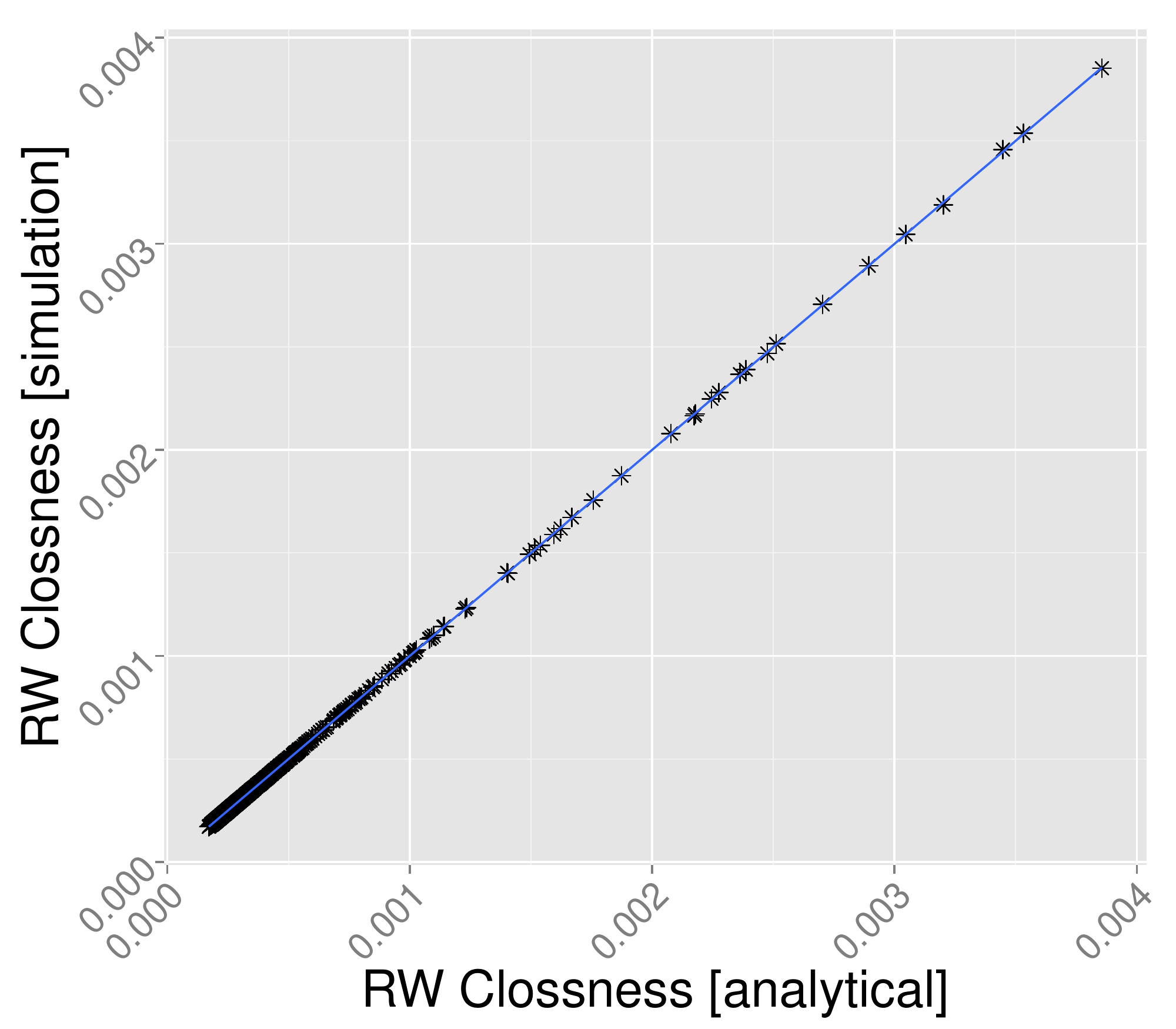}
        \end{tabular}
    \end{center}
        \caption{Comparison of the random walk closeness centrality obtained by simulation and by our analytical approach for different multilayer network topologies. Each multilayer network is composed of two layers with 1000 nodes per layer. {\bf A}, results on a multilayer network with two Erd{\H{o}}s-R\'{e}nyi networks as layers. {\bf B}, results on a multilayer network with one Erd{\H{o}}s-R\'{e}nyi network and one Barab\`{a}si-Albert network as layers. {\bf C}, results on a multilayer network with two Barab\`{a}si-Albert networks as layers.}
        \label{fig:evaluationCloseness}
\end{figure*}

The average mean first passage time $h_{[d]}$ to node $d$ is obtained by averaging $\(H_{[d]}\)^{o\sigma}$ over all possible starting nodes and layers as
\begin{eqnarray}
\label{eq:defMeanTime}
	h_{[d]} = \frac{1}{(N-1)L}u_{o\sigma}\(H_{[d]}\)^{o\sigma} + \frac{1}{N}\pi^{-1}_{[d]},
\end{eqnarray}
where $\pi_{[d]}$ is the occupation probability of node $d$ and the term $\frac{1}{N}\pi^{-1}_{[d]}$ is included explicitly to account for the average return time, that is not accounted for when using absorbing random walks.

Finally, the random walk closeness centrality of node $d$ is defined as the inverse of $h_{[d]}$. We introduce the vector $\xi_{i}$ whose components are given by the inverse of the corresponding values of $h$.

The comparison between the values of $\xi_{i}$ obtained from simulations and theoretical predictions are shown in Fig.\,\ref{fig:evaluationCloseness}. As in the betweenness centrality, the results are in excellent agreement and it is worth remarking that the equivalence holds regardless of the number of nodes in the network, the topology and the number of layers.


\section{Conclusions and Discussion}\label{sec:conclusion}

We have extended the main random walk centrality measures to interconnected multilayer networks and gave interpretation of their meaning. In addition, we have presented analytical approaches, based on the tensorial formalism defined in \cite{dedomenico2013mathematical}, for their computation. The comparison of the predictions given by our analytical approach with the results obtained by simulations show a perfect agreement, concluding that the presented analytical expressions are ready to be applied to the analysis of real complex networks. We expect that the presented results are useful in many interdisciplinary applications ranging from social sciences to transportation networks.


\section*{Acknowledgements}

AA, MDD, SG, and AS were supported by the European Commission FET-Proactive project PLEXMATH (Grant No. 317614), the MULTIPLEX (grant 317532) and the Generalitat de Catalunya 2009-SGR-838. AA also acknowledges financial support from the ICREA Academia and the James S.\ McDonnell Foundation, and SG and AA were supported by FIS2012-38266.


\appendix


\section{Eigenvalue problem with tensors}\label{sec:eigentensor}

The eigenvalue problem for a rank-2 tensor, i.e., a standard matrix, is defined by $W^{i}_{j}v_{i}=\lambda v_{j}$. The extension of this problem to rank-4 tensors leads to the equation
\begin{eqnarray}
M^{i\alpha}_{j\beta}V_{i\alpha}=\lambda V_{j\beta}.
\end{eqnarray}
To solve this problem, it is worth noting that any tensor can be \emph{unfolded} to lower rank tensors \cite{kolda2009}. For instance, a rank-2 tensor like $W^{i}_{j}$, with $N^{2}$ components, can be flattened to a vector $w_{k}$ with $N^{2}$ components. In the case of the rank-4 multilayer adjacency tensor $M^{i\alpha}_{j\beta}$, although any unfolding is allowed, it is particularly useful for some applications to choose the ones flattening to a squared rank-2 tensor $\tilde{M}^{k}_{l}$ with $NL\times NL$ components, where $L$ indicates the number of layers \cite{gomez2013diffusion}. In fact, this unfolding produces as many block adjacency matrices, named \emph{supra-adjacency matrices} in some applications \cite{gomez2013diffusion,dedomenico2013random,cozzo2013cc}, as the number of permutations of diagonal blocks of size $N^{2}$, i.e., $L!$. However, such unfoldings do not alter the spectral properties of the resulting supra-matrix and can be used to solve the eigenvalue problem for rank-4 tensors. In fact, the solution of the eigenvalue problem
\begin{eqnarray}
\tilde{M}^{k}_{l}\tilde{v}_{k}=\tilde{\lambda}_{1}\tilde{v}_{l},
\end{eqnarray}
is a \emph{supra-vector} with $NL$ components which corresponds to the unfolding of the eigentensor $V_{i\alpha}$.


\section{Mean number of crossing times} \label{ap:timesOnNodeRWB}

Given $M$ random walks starting in node $o$ on layer $\sigma$ and ending when reaching node $d$, regardless of the layer, the expected number of times a random walk will pass by node $j$ on layer $\beta$ is given by
\begin{eqnarray}\label{eq:meanNumberOftimes}
\(\mathcal{T}_{[d]}\)^{o\sigma}_{j\beta} &=& \lim_{M \to \infty} \frac{1}{M} \sum\limits_{m=1}^{M} \sum\limits_{t=0}^{\infty} z_{j\beta}^{o\sigma}(t,m),
\end{eqnarray}
where $z_{j\beta}^{o\sigma}(t,m)=1$ if walk $m$ was visiting node $j$ in layer $\beta$ at time step $t$ and $z_{j\beta}^{o\sigma}(t,m)=0$ otherwise.

Following the frequentist interpretation, the probability of being in node $j$ in layer $\beta$ at time step $t$, provided that the walk originated in node $o$ in layer $\sigma$, is given by
\begin{eqnarray}\label{eq:probOfBeingAsLimit}
p_{j{\beta}}^{o{\sigma}}(t) = \lim_{M \to \infty} \frac{1}{M} \sum\limits_{m=1}^{m} z_{j\beta}^{o\sigma}(t,m).
\end{eqnarray}
Substituting (\ref{eq:probOfBeingAsLimit}) in (\ref{eq:meanNumberOftimes}) we obtain that
\begin{eqnarray}
\(\tau_{[d]}\)^{o\sigma}_{j\beta} &=& \sum\limits_{t=0}^{\infty} p_{j{\beta}}^{o{\sigma}}(t) = \sum\limits_{t=0}^{\infty} \(T_{[d]}^{t}\)^{o\sigma}_{j\beta} \nonumber\\
&=& \[\(\delta - T_{[d]}\)^{-1}\]^{o\sigma}_{j\beta}
\end{eqnarray}
where $T_{[d]}$ corresponds to the absorbing transition tensor defined in Eq.\,(\ref{eq:absTranTensor}).

\end{document}